\documentclass[%
reprint,
superscriptaddress,
amsmath,amssymb,
pra,
floatfix,
]{revtex4-1}
\usepackage{url}
\usepackage{graphicx}
\usepackage{upgreek}

\begin{document}

\title{Cryogenic Electro-Optic Polarisation Conversion in Titanium in-diffused Lithium Niobate Waveguides}




\author{Frederik Thiele}
\affiliation{ 
Mesoscopic Quantum Optics, Department Physics, Paderborn  University, 33098 Paderborn Warburger Str. 100, Germany
}%
\author{Felix vom Bruch} \author{Victor Quiring} \author{Raimund Ricken} \author{Harald Herrmann} \author{Christof Eigner} \author{Christine Silberhorn} 
\affiliation{ 
Integrated Quantum Optics, Department Physics, Paderborn  University, 33098 Paderborn Warburger Str. 100, Germany
}%
\author{Tim J. Bartley}%
\affiliation{ 
Mesoscopic Quantum Optics, Department Physics, Paderborn  University, 33098 Paderborn Warburger Str. 100, Germany
}%






\begin{abstract}
    Many technologies in quantum photonics require cryogenic conditions to operate. However, the underlying platform behind active components such as switches, modulators and phase shifters must be compatible with these operating conditions. To address this, we demonstrate an electro-optic polarisation converter for 1550\,nm light at 0.8\,K in titanium in-diffused lithium niobate waveguides. To do so, we exploit the electro-optic properties of lithium niobate to convert between orthogonal polarisation modes with a fiber-to-fiber transmission $>$43\%.  We achieve a modulation depth of 23.6$\pm$3.3\,dB and a conversion voltage-length product of 28.8\,V\,cm. This enables the combination of cryogenic photonics and active components on a single integration platform.
\end{abstract}

\maketitle
\section{Introduction}
    The current state-of-the-art in high-speed single photon detectors are based on superconducting nanowires \cite{Marsili2013,Hopker2019}; such detectors have demonstrated timing jitters below 5ps \cite{Korzh}. Modulation in quantum photonics typically exploits advances in integrated optics, although the lowest loss implementations still rely on bulk devices to maximise efficiency. This makes integration with superconducting detectors extremely challenging due to the cryogenic operating requirements. Indeed, many modulation techniques in integrated platforms also rely on approaches which are incompatible with cryogenic or low-loss operation, such as thermo-optic\cite{Harris2014} or carrier-injection\cite{Gehl2016}.  

    By contrast, integrated optics based on titanium in-diffused waveguides in lithium niobate platform offers several advantages in this regard. As a material with a high electro-optic coefficient, it is compatible in principle with cryogenic operation. Furthermore, its large mode sizes are well-matched to optical fibre, which reduces overall losses. Indeed, many quantum photonic components have been integrated on this platform. However, quantum-compatible operation of these components under cryogenic conditions - and thereby compatibility with integrated superconducting detectors, remains an open challenge.
    
    While phaseshifting has been achieved at cryogenic temperatures in lithium niobate~\cite{CryoPhase,McConaghy1996,Herzog2008,Youssefi2020} and other platforms~\cite{Gehl2016,Eltes2019}, low-temperature operation of active polarisation conversion is yet to be demonstrated. The polarisation degree of freedom is a very natural basis for encoding qubits on a single photon level, therefore efficient control is an important tool in many quantum photonic information tasks. Moreover at room temperature, active polarisation conversion has been established since the 1980s~\cite{Alferness1980,Thaniyavarn1985,Lu2000,Huang2007b,Moeini2010} and has also been developed in thin-film lithium niobate (LNOI)~\cite{Izuhara2003,Ding2019a}. However, until now, active polarisation conversion at low temperatures has not been demonstrated. 
    
    In this paper, we demonstrate that polarisation conversion in z-cut Ti:LN is indeed fully functional at the temperatures required for superconducting detectors. We show electro-optic conversion between orthogonal polarisation modes from room temperature down to 0.8\,K. This expands the degrees of freedom with which single photons can be manipulated for quantum photonic tasks~\cite{Wang2019a} whilst maintaining compatibility with the operating conditions of  superconducting detectors, as well as other low-temperature quantum photonic technologies such as many single photon emitters.
    
    This paper is organised as follows. We first describe the operation of the electro-optic polarisation converter in terms of a temperature-dependent quasi-phase-matched nonlinear process. We then present the experimental methods for testing the device in a robust, plug-and-play configuration. The results are presented in section~\ref{sec:Results} before summarising in section~\ref{sec:Concl}.

\section{Temperature Dependence of  Polarisation Conversion}
    
    \subsection{Phase-Matching}\label{sec:Phase-matching}
        \begin{figure*}
            \centering
            \includegraphics[width=1.0\linewidth]{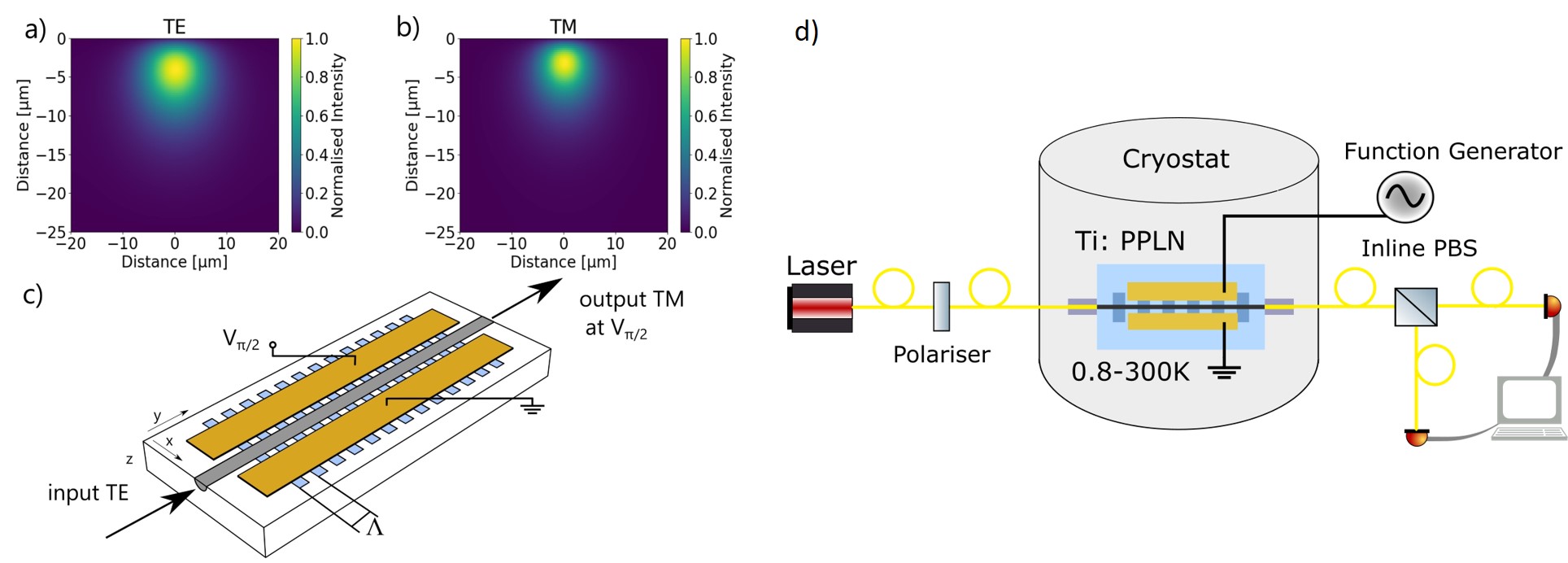}
            \caption{a), b) The normalised spatial intensity distribution of the supported optical modes, TE and TM, respectively. c) Schematic of the polarisation converter, including poling period $\Lambda$. The linear polarisation mode TE is converted to TM if the voltage of $V_{\pi/2}$ is applied at the electrodes. d) Schematic of the optical characterisation set up. All components other than the waveguide operate at room temperature. The sample is optically accessed via fiber feed-throughs.}
            \label{fig:modulator}
        \end{figure*}
        Polarisation conversion in lithium niobate is a phase-matched interaction between two orthogonally polarised optical modes~\cite{Yariv} (henceforth referred to as transverse electric - TE, and transverse magnetic - TM) and a static electric field. The interaction can be modeled as a coupling between the TE and TM modes, mediated by the DC electric field $E_\textrm{DC}$, in a configuration shown in Fig.~\ref{fig:modulator}. The strength of the coupling depends on the crystal orientation of the lithium niobate and the corresponding electro-optic coefficient, as well as the spatial mode overlap $\eta = \int_A E_\textrm{TE}(x,z) E_\textrm{TM}(x,z) E_\textrm{DC}(x,z) \mathrm{d} A$ of the polarisation modes $E_\textrm{TE},~E_\textrm{TM}$ and electric field $E_\textrm{DC}$. 
        
        An important property of the waveguides in Ti:LN is that they support low-loss propagation of orthogonal polarisation modes with a high degree of overlap (as shown in Fig.~\ref{fig:modulator}~a)); however, since lithium niobate is a birefringent material, the propagation constants for each mode are different. However, a periodic domain inversion in the crystal, with a chosen period length $\Lambda$, can be used to compensate the phase mismatch between the two polarisation modes at a specific wavelength $\lambda_\textrm{pm}$. Given temperature- and wavelength-dependent refractive indices $n_\textrm{TE}\left(\lambda,T\right),~n_\textrm{TM}\left(\lambda,T\right)$, the phase mismatch between the modes $\Delta\beta$ is given by
                    \begin{equation}
                        	\Delta \beta = 2 \pi \left(\frac{n_\textrm{TE}(\lambda,T)-n_\textrm{TM}(\lambda,T)}{\lambda}\right)-\left(\frac{2 \pi}{\Lambda}\right)~.
                    \end{equation}
    
        While the period length $\Lambda$ remains broadly temperature insensitive, it is clear that the phase-matched wavelength $\lambda_\textrm{pm}=\Lambda\left(n_\textrm{TE}(\lambda,T)-n_\textrm{TM}(\lambda,T)\right)$, for which $\Delta\beta=0$, will vary with temperature.
        
\begin{figure}
                    \centering
                    \includegraphics[width=1\linewidth]{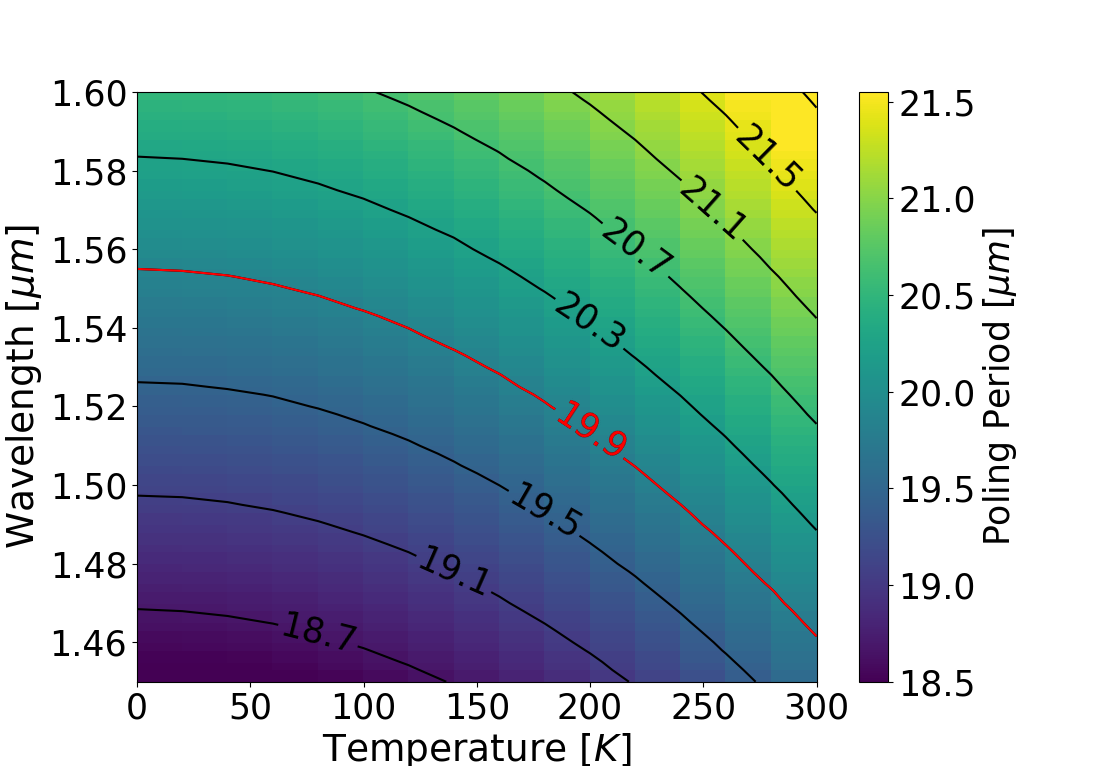}
                    \caption{Extrapolated poling period from the Sellmeier equations in combination with the titanium in-diffusion profile required to phase-match the conversion process at a given wavelength and temperature.}
                    \label{fig:birf2d}
                \end{figure}

        To phase-match the conversion process in the optical telecommunication band at room temperature, a poling period of around 20$\,\upmu$m is required. By extrapolating the Sellmeier equations for the refractive indices for our waveguide system~\cite{PropLN} down to cryogenic temperatures around 1\,K, and taking into account the effective refractive index caused by the titanium diffusion profile of our waveguides, we expect a shift in the phase-matched wavelength by up to 100\,nm. This is indicated in Fig.~\ref{fig:birf2d}. We thus choose a poling period of 19.9\,$\upmu$m, which is expected to be phase-matched around 1560.5\,nm at cryogenic temperatures and 1467.4\,nm at room temperature.
        Phase-matching across this range of wavelengths can be tested using a tunable laser of suitable tuning range, as described in Section~\ref{sec:Method}.

\subsection{Coupled Mode Equations }\label{sec:Coupled}

    Given initial amplitudes of the orthogonal modes $A_\textrm{TE}\left(0\right),~A_\textrm{TM}\left(0\right)$, the solution of the coupled mode equations describing the conversion process after propagating a distance $y$ are given by~\cite{Yariv}
                \begin{eqnarray}
                                \begin{pmatrix}   A_\textrm{TE}\left(y\right) \\ A_\textrm{TM}\left(y\right) \end{pmatrix} = \mathrm{M}_y \cdot e^{\frac{i \Delta \beta y}{2}} \begin{pmatrix}   A_\textrm{TE}\left(0\right) \\ A_\textrm{TM}\left(0\right) \end{pmatrix}~,
                \end{eqnarray}
     where $\mathrm{M}_y$ is the transfer matrix
                 \begin{equation}
                     \mathrm{M}_{y} = \begin{pmatrix} \cos(sy)+ \frac{i\Delta \beta}{2 s} \sin(sy)	 & -  \frac{\kappa}{s} \sin(sy)\\ +  \frac{\kappa}{s} \sin(sy) & \cos(sy)- \frac{i\Delta \beta}{2 s} \sin(sy) \end{pmatrix}~,
                 \end{equation}
    with $s=\sqrt{\kappa^2+\left({\Delta \beta}/{2}\right)^2}$ and the coupling strength $\kappa$ given by 
                \begin{equation}
                \kappa = \frac{2\pi \,n_\textrm{eff}^3 \eta r V}{\lambda}~,
                \end{equation}
    where $V$ is the applied bias voltage and $r$ is the electro-optic coefficient dependent on the direction of the applied field. The effective refractive index is here defined as $n_\textrm{eff}^{3} =n_{TE}^{3/2}n_{TM}^{3/2} $ and the spatial overlap $\eta $ is determined by the spatial overlap of the electric fields.

\section{Experimental Implementation}\label{sec:Method}
    To demonstrate the effects of temperature on the electro-optic conversion process, we fabricated and characterised a robust, fibre-coupled device based on titanium in-diffused waveguides in periodically-poled, z-cut lithium niobate~\cite{Sohler2008a,Sharapova2017}. Despite the lower electro-optic coefficient of z-cut lithium niobate, this cut was chosen in order to be forward compatible with other nonlinear processes such as frequency conversion. 
  
      \subsection{Device Geometry}
          The device itself consists of a waveguide of length 25\,mm, which is is poled with a period of 19.9\,$\upmu$m. The poling structures are fabricated using liquid electrodes, followed by application of high voltage pulses to invert the crystal. After the poling procedure, gold electrodes with a buffer-layer of 200\,nm $\mathrm{SiO_2}$ in between were deposited, with a length of 15\,mm, a separation of 15\,$\upmu$m and a width of 400\,$\upmu$m. This electrode structure should act as an ideal capacitor, of capacitance of 9.2\,pF, given the electrode geometry~\cite{Alferness1982}. As such, the converter should not dissipate any electric energy, which would be detrimental to operation inside a low-temperature cryostat. In practise, however, impedance miss-matches and stray resistances throughout the circuit may cause some electrical energy dissipation. Wire bonds from the electrodes to electrical connectors allow an external bias voltage to be applied via a coaxial cable. The linear losses in these waveguides are measured to be about $0.17\,\mathrm{dB/cm}$ in the TE-polarisation at room temperature in the optical C-band \cite{Regener1985}.

     \subsection{Packaging}    
         To enable operation of the sample in a cryostat, the device was butt-coupled (``pigtailed'') to single mode polarisation maintaining (PM) fibres using UV-curable adhesive (Norland 81). Together with an anti-reflective coating between the fibre and sample, fiber-to-fiber efficiency of $55\,\%$ were achieved at room temperature. For mechanical robustness during temperature cycling, the sample is mounted on a copper holder and the fibres are attached to additional side blocks for strain relief. These fibres are then spliced (with approximately 0.01dB additional loss per splice) to PM fiber feed-throughs, providing optical access in and out of the cryostat. When cooled to cryogenic temperatures, coupling efficiencies of 43\% were achieved. Changes in the coupling efficiency can be attributed to thermal stresses at the butt-coupled joints. 
         
     \subsection{Characterisation Methods}     
\begin{figure*}
                \centering
                \includegraphics[width=1.0\linewidth]{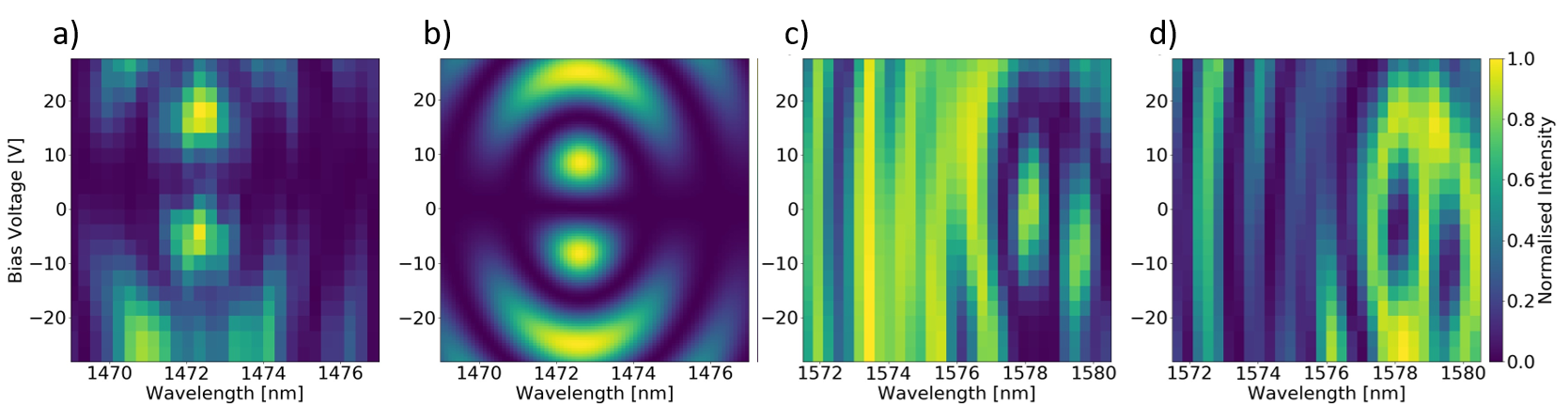}
                \caption{Conversion maps showing the normalised intensity of the TM polarisation mode by modulating the bias voltage and wavelength given a TE input polarisation. a) Experimental data of conversion at room temperature. b) Theoretical conversion map determined by the transfer matrix in Sec.~ \ref{sec:Coupled}. Conversion maps for c) TE and d) TM output polarisation at 0.8\,K given a TE input polarisation.}.
                \label{fig:PlotFigures}
            \end{figure*}

        To characterise the device at cryogenic temperatures, the device is mounted on a cold stage of a closed-cycle cryostat (PhotonSpot Cryospot 4) which can reach a base temperature of 0.77\,K, as shown in Fig.~\ref{fig:modulator}~d).
        During the device cooling process, light of one polarisation was coupled into the device, and the output polarisation was optically monitored. Effects of both wavelength and applied voltage were monitored during the cooling process. Using a tunable laser (EXFO T100S-HP/SCL) with a power of approximately 200\,$\upmu$W, the wavelength was swept in 0.3\,nm steps in a range of 9\,nm centred at a wavelength of 1472.5\,nm, and the bias voltage swept from -30 to 30\,V in 2.5\,V steps. The entire sweep lasts approximately 100 seconds. The resulting intensity modulation at 296\,K is represented as a conversion map in Fig.~\ref{fig:PlotFigures}~a) and is in a good agreement with the theoretically determined conversion map in Fig.~\ref{fig:PlotFigures}~b), determined by the derived transfer matrix in Sec.~\ref{sec:Phase-matching}. 
        
        This sweep was repeated during temperature cycling, whereby the center wavelength around which the laser was swept was based on a tracking algorithm which sought the wavelength which maximised conversion efficiency. Using this technique, 576 wavelength-voltage conversion maps, such as Fig.~\ref{fig:PlotFigures}~a) and b), were acquired across a temperature range from 296-0.8\,K, and an overall wavelength range of 1468\,nm to 1581\,nm. 
        The overall cool-down time is around 16 hours, with a maximum cooling rate of approximately 0.35\,K per minute. This method results in a temperature uncertainty across the switching maps of less than 1\,K.
    
\section{Results}\label{sec:Results}
    The primary result is that polarisation conversion was observed down to 0.8\,K, at a wavelength of 1578.2\,nm, with a modulation depth of 23.6$\pm$3.3\,dB, and a modulation voltage of 19.1$\pm$2.1\,V. The switching map at this temperature, as shown in Fig.~\ref{fig:PlotFigures}~c) and d), shows significant deviations from the expected behavior, and quantitative analysis of this discrepancy is highly challenging. We suggest that it is the effects of pyro-electric charge accumulation due to the temperature changes which play a significant role in affecting the field distribution both inside and outside the poled region of the waveguide, giving rise to additional polarisation mode coupling conditions. Despite these effects, by considering solely the phase-matched wavelength, the device retains its functionality across the whole temperature range. 
    In terms of cryogenic compatibility, optical losses are the primary source of thermal heat load of the device. When testing at base temperature, the 43\% fiber-to-fiber efficiency and testing at an optical power level of ca. 200\,$\upmu$W raised the base temperature by around 200\,mK to 1\,K (over a timescale of ca. 10 minutes). This base temperature is well-within the operating requirements of superconducting detectors, and furthermore the temperature increase would not be present if the device was operating on the single-photon regime. Addition thermal effects arising from the modulation circuitry were not observed, and would anyway depend on the operating speed of the device.
 
    Using the conversion maps obtained during temperature cycling (see Visualisation 1 \cite{Thiele2020b}), we characterise our device at the phase-matched wavelength over the full temperature range.
    We define the phase-matched wavelength to be the wavelength at which the modulation depth is maximised, (whereby the modulation depth is defined at the ratio of maximum to minimum intensity when the voltage is swept).
     

    \subsection{Phase-Matched Wavelength}   
    
\begin{figure*}
                \centering
                \includegraphics[width=1.00\linewidth]{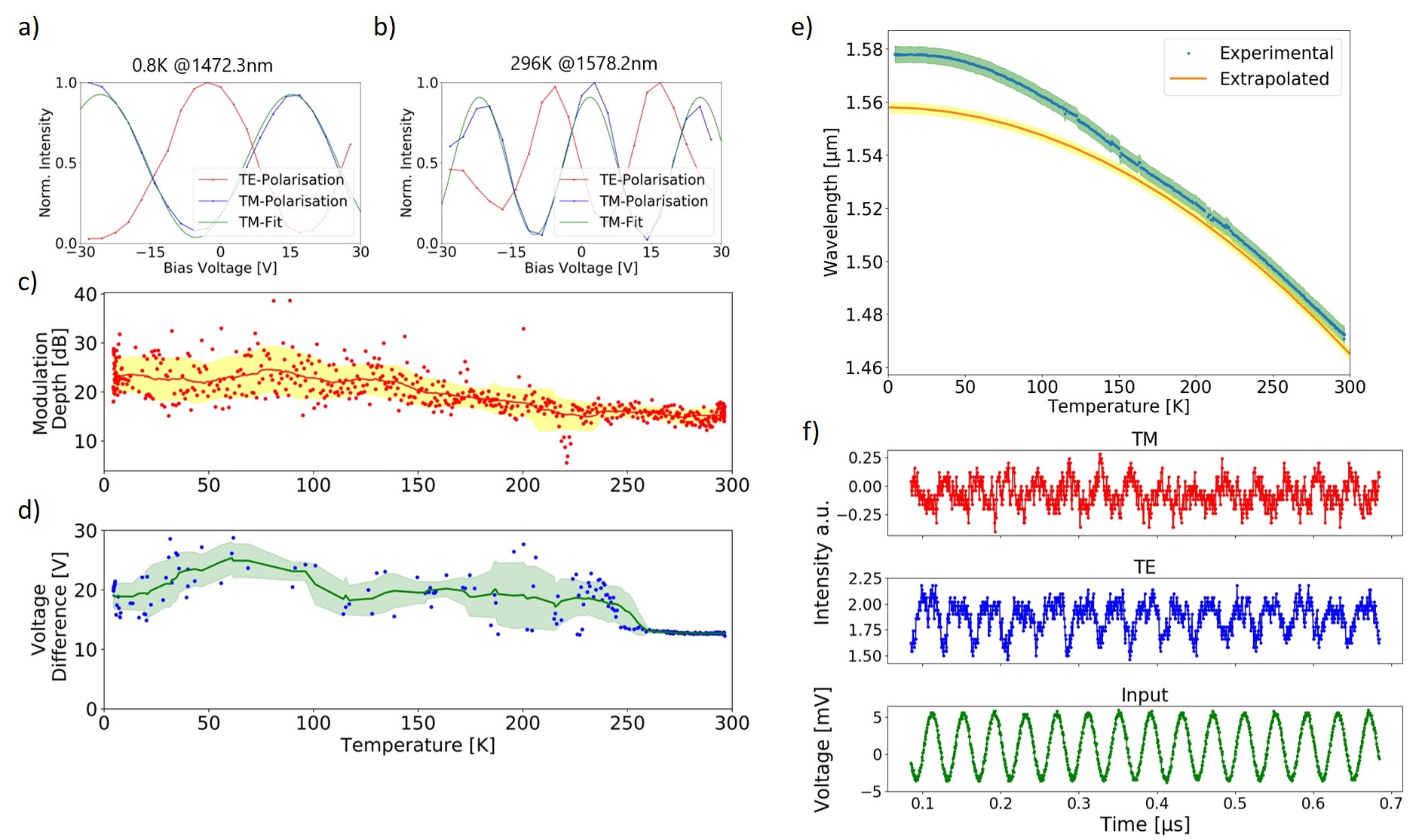}
                \caption{a), b): Voltage dependent intensity modulation at 0.8\,K for 1578.2\,nm and at 296\,K for 1472.3\,nm, respectively. c): Maximum modulation depth determined across the temperature range. The modulation depth is the ratio of the minimum and maximum intensity in the voltage sweeps, {e.g.} a) and b). The yellow band indicates a 15\,K moving average, corresponding to . d): Conversion voltage $V_{\pi/2}$ required to convert TE to TM, as a function of temperature. This can be determined by the period of the sine-fit as seen in the voltage sweeps in a), b). The green band is the $V_{\pi/2}$-average in the temperature range of 10\,K. e) Phase-matched wavelength as a function of temperature. Blue dots \& green region: experimental data and experimental error, respectively. Red line \& yellow region: extrapolated wavelength and simulation uncertainty, respectively. f) Time-dependent modulation of the polarisation converter at 0.8\,K at 25\,MHz. The device is biased by 9\,V to be operated at the point of inflection of the voltage sweep.}
                \label{fig:SrVoltage}
            \end{figure*}

        The dependence of the phase-matched wavelength on temperature is shown in Fig.~\ref{fig:SrVoltage}~e). Operation across the whole temperature range was maintained, and as expected, the phase-matched wavelength increases with decreasing temperature. However, the magnitude of the shift, namely 107\,nm, is larger than the 94\,nm predicted by extrapolating the Sellmeier equations. We also observe a small quantitative offset between the model and the measured value at room temperature, which we attribute to imperfections of the modelling of the titanium diffusion profile of the waveguide. Overall, the extrapolation agrees reasonably well with the phase-matched wavelength down to 150\,K, but deviates significantly at lower temperatures. This shows that further corrections to the Sellmeier equations in this temperature range are required. The experimental data also matches the qualitative features of the simulation, in that the dependence of the center wavelength with temperature approaches zero as the temperature approaches 0\,K. 
        
                            
        
    \subsection{Modulation Depth}
        In Fig.~\ref{fig:SrVoltage}~a) and b) we show how the power in each polarisation mode depends on applied voltage, carried out at a fixed temperature and wavelength. From these plots, we determine the modulation depth as the ratio of the maximum to minimum intensity. Fig.~\ref{fig:SrVoltage}~c) shows how the modulation depth changes as a function of temperature. Throughout the entire temperature range, a modulation depth of the device was maintained around 15-25\,dB, even increasing slightly at lower temperatures. At 0.8\,K, we measure a modulation depth of 23.6$\pm$3.3\,dB, with error bars determined statistically. This demonstrates that polarisation can be manipulated with the precision required for quantum optical experiments, {i.e.} that polarisation states with a fidelity of 99.6\% can be reached.  
        

    \subsection{Modulation Voltage}
        From Figs.~\ref{fig:SrVoltage}~a) and b), we determine the modulation voltage at the phase-matched wavelength by the period of a sinusoidal fit to the voltage sweep. The modulation voltage required for a full polarisation conversion is in the range of 12.7$\pm$0.1\,V at room temperature for the 15\,mm long electrodes. This increases to 19.2$\pm$2.1\,V at 0.8\,K during the cooling process as shown in Fig.~\ref{fig:SrVoltage}~d), which results in a voltage-length switching product of 28.8\,V\,cm. Further optimisations to electrode geometries would depend on the required application; high-speed applications typically require low conversion voltages while fine polarisation control is achieved with high conversion voltage-length products. 
        
        It is also apparent from Fig.~\ref{fig:SrVoltage}~d) that the switching voltage increases and becomes much more noisy in the temperature range 10\,K-250\,K, which we again attribute to the accumulation of pyro-electric charge, since the rate of temperature change is maximal in this area. This changes the conversion properties at no externally applied voltage (0\,V), but should be relatively straightforward to compensate by an additional D.C. bias voltage.   

    \subsection{Modulation Speed}
        For application in a quantum circuit, the operating speed of active components is an important parameter. In principle, the electrode design of our polarisation converter permits fast electro-optic control~\cite{Alferness1982}. In practice, operating speeds are limited by driving electronics as much as the active devices themselves. Moreover, operating inside a low-temperature cryostat adds additional complexity. Nevertheless, in a proof-of-principle experiment to investigate typical operating speeds, we tested our device at 0.8\,K when connected to a 25\,MHz function generator. This created a sine-signal with an amplitude of 10\,mV, which was limited by the high impedance of the electrodes. The voltage signal was biased by about 9\,V so that the modulation is incident at the point of inflexion of the voltage sweep. The output signal was tracked with an NIR-diodes and an oscilloscope. Plots showing the response of the diodes to each polarisation are shown in Fig.~\ref{fig:SrVoltage}~f). A modulation depth of around 1.8\,dB was achieved, which is strongly limited by the signal-to-noise ratio of the diodes and maximal applied voltage amplitude on the electrode structure. Although a complete crossover in the polarisation was not reached, this experiment shows that MHz modulation is still possible in principle under cryogenic conditions. 

\section{Conclusion}\label{sec:Concl}
In conclusion, we have presented an electro-optic polarisation converter at 0.8\,K in titanium in-diffused lithium niobate waveguides. In so doing, we have demonstrated mutual compatibility between active electro-optic components for quantum photonic circuits and the operating conditions required for superconducting detectors. Under these conditions, we have shown a fibre-to-fibre coupling efficiency of 43\% and a modulation depth of 23.6$\pm$3.3\,dB. 
    
    When observing the dynamics during the cooling process, the change in temperature shifts the phase-matched wavelength by 107\,nm, which is larger than predicted by extrapolating existing Sellmeier equations in combination with our waveguide modelling. Furthermore, we observe an increase in the  modulation voltage from 12.7$\pm$0.1\,V to 19.2$\pm$2.1\,V, which suggests temperature dependence of the electro-optic coefficient in lithium niobate. The polarisation converter is capable of modulation at 25\,MHz at a temperature of 0.8\,K. Operation was preserved across the whole temperature range, despite complex dynamics during the cooling process which we attribute to pyro-electric charge accumulation. However, the device operates in a stable manner once at cryogenic temperatures. 
    This functionality shows that polarisation conversion is compatible with other low-temperature technologies required for integrated quantum photonics.

\section*{Funding}
This project is supported by the German Federal Ministry of Education and Research (BMBF) under the funding program Photonics Research Germany, grant number 13N14911;
The Integrated Quantum Optics Group acknowledges funding from the European Commission through the ERC project QuPoPCoRN (Grant No. 725366).

\section*{Disclosures}
The authors declare no conflicts of interest.


\bibliography{Main}






\end{document}